\documentclass[letterpaper,12pt]{article}   %% LaTeX 2e (preferred)
\usepackage{osajnl2} %% do not use with REVTeX4
\usepackage{comment}
\usepackage{graphicx}
\usepackage{subfig}

\begin{document}

\title{Phase estimation with two-mode squeezed-vacuum and parity detection restricted to finite resources}

\author{Keith R.\ Motes,$^{1,*}$ Petr M.\ Anisimov,$^1$ and Jonathan P.\ Dowling$^{1,2}$}
\address{$^1$ Hearne Institute for Theoretical Physics and Department of Physics and Astronomy \\
Louisiana State University, Baton Rouge, LA 70803}
\address{$^2$ Computational Science Research Center, Beijing 100084, China}
\address{$^*$Corresponding author: motesk@gmail.com}

\begin{abstract}
A recently proposed phase-estimation protocol that is based on measuring the 
parity of a two-mode squeezed-vacuum state at the output of a Mach-Zehnder
interferometer shows that Cram\'{e}r-Rao bound sensitivity can be obtained 
[P.\ M.\ Anisimov, et al., Phys.\ Rev.\ Lett.\ {\bf104}, 103602 (2010)]. This 
sensitivity, however, is expected in the case of an infinite number of parity 
measurements made on an infinite number of photons. Here we consider the case of a finite number of parity measurements and a finite number of photons, implemented with photon-number-resolving detectors. We use Bayesian
analysis to characterize the sensitivity of the phase estimation in this scheme. We have found that our phase estimation becomes biased near $0$ or $\pi/2$ 
phase values. Yet there is an in-between region where the bias becomes 
negligible. In this region, our phase estimation scheme saturates the
Cram\'{e}r-Rao bound and beats the shot-noise limit. 
\end{abstract}
%Pick some here
%270.0270   Quantum optics, 
%270.5570   Quantum detectors, 
%270.6570   Squeezed states, 
%120.3180   Interferometry, 
%120.3940   Metrology, 
%120.5050   Phase measurement%

\maketitle %% null function with osajnl.sty

\section{Introduction}
%============================================================================
Phase estimation is a primary objective of optical quantum metrology. There are several experimental ways to estimate phase. Coherent light based interferometry is most commonly used but its sensitivity for phase estimation is limited by the shot-noise (SN) limit, $(\Delta \theta)^{2}\ge \bar{n}^{-1}$ \cite{Dowling2008}. This is not a problem in the case of limitless resources or in the case of samples that can withstand large doses of radiation. However, this is a problem otherwise, and one has to resort to interferometry with a finite number of quantum states of light, such as N00N states \cite{LeeDowling2002}, and measuring parity \cite{a:Gerry.2010, PhysRevA.68.023810} in order to achieve
sub-shot-noise or even Heisenberg-limited (HL) sensitivity of phase estimation. 

%Christopher Gerry and collaborators first introduced the parity operator in the context of quantum metrology in a %seminal paper in 2003 \cite{PhysRevA.68.023810}. The operator is also covered in the text book by Gerry and Knight %\cite{gerry2005introductory}. For a splendid and exhaustive review of the use of this operator please peruse the %sublime article by Gerry and Jihane in Ref.\cite{doi:10.1080/00107514.2010.509995}.

\begin{figure}
\centering
\subfloat[Part 1][Mach-Zehnder interferometer (MZI)]
{\includegraphics[width=2.5in]{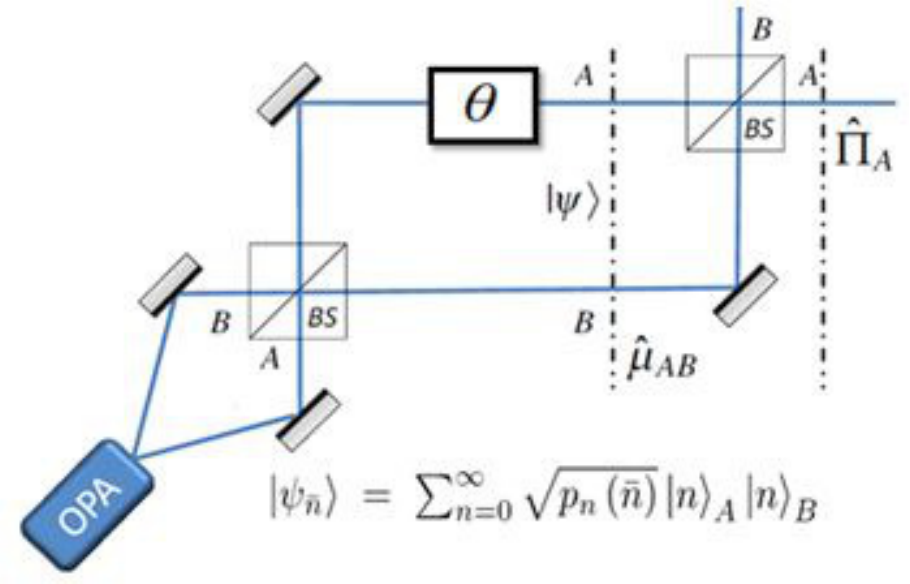} \label{fig:MZI}}
\subfloat[Part 2][Convergence of Parity Signal]
{\includegraphics[width=2.5in]{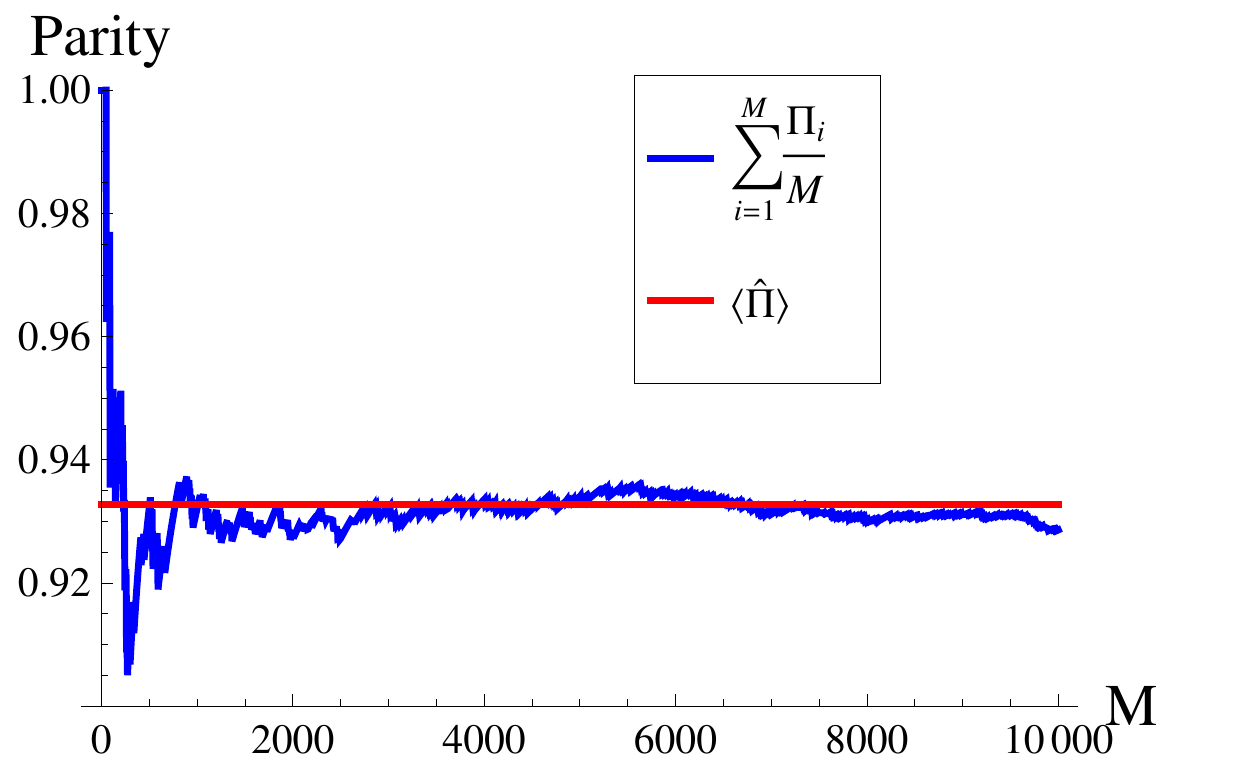} \label{fig:ParityBuildUp}}
\caption{(a) Two-mode squeezed-vacuum states are generated at the
input of the MZI by an optical parametric amplifier. The parity
signal at the output of the MZI can be measured with a photon-number-resolving detector.
(b) Convergence of the parity signal, based
on outcomes of numerically generated detection events (blue), to the expected
value of $\langle\hat{\Pi}\rangle=0.93$ (red), in a finite number of $M=10^4$
parity measurements. A different sample of this convergence will evolve
differently due to the probabilistic nature of the scheme, but we find it
always converges to the expectation value. Here, we assumed that the photon
number is $\bar{n}=3$ and the unknown phase had value $\theta= 0.1$.}
\label{fig:FigureOne}
\end{figure}

Significant advances have been made in quantum-enhanced phase
sensitivity \cite{a:Lloyd.2011} and the meaning of the Heisenberg limit has been
thoroughly examined \cite{a:Zwierz.2012, bib:PhysRevA.85.041802}. Yet, a recently proposed phase
estimation scheme dips below the HL in the case of an infinite number of parity
measurements \cite{PhysRevLett.104.103602}. That scheme is based on measuring
the parity of the state of light at the output of a Mach-Zehnder
interferometer (MZI), as shown in Fig.~\ref{fig:MZI}, with two-mode squeezed-vacuum (TMSV)
input. It turned out that this particular scheme using TMSV input has
sub-Heisenberg sensitivity even with linear phase evolution. This is due to
the fact that the photon number uncertainty for the state of light inside of the
MZI is greater than the average photon number used for the measurement
\cite{PhysRevA.79.033822,PhysRevLett.105.120501}.

In this paper we define the Heisenberg limit, following the usual convention, to be $\Delta\theta \equiv 1/\bar{n}$, where $\bar{n}$ is the \textit{average} number of photons \cite{Dowling2008}. The term \textit{Heisenberg limit} was \textit{defined} by Holland and Sanders to be $\Delta\theta\equiv1/N$ for states with $N$ fixed total number of photons such as twin-Fock or N00N states \cite{bib:holland1993interferometric}. This limit is a rigorous lower limit for \textit{local} phase sensitivity for such states \cite{PhysRevLett.99.070801}. However, for states with well-defined mean photon number but undefined total photon number, such as the TMSV used here, it is now understood that the HL so defined is not a hard lower limit. 

Several recent papers have generalised the definition of the HL so that even for such states of ill-defined photon number the new limit cannot be exceeded \cite{bib:PhysRevA.86.053813, bib:zwierz2012ultimate}. However, to avoid confusion, given that there are now several different definitions of the HL in the literature, we concern ourselves primarily here with the quantum Cram\'{e}r-Rao bound, which is provably the ultimate limit of phase sensitivity. Hence if we saturate this limit, as we do here with parity detection, there can be no ambiguity in the interpretation, and our results in this sense can be said to be optimal. 

We continue to use the conventional definition, following Holland and Burnett, that $\Delta\theta \equiv 1/\bar{n}$ is \textit{the} Heisenberg limit and instead we refer to all these other limits that have the form $\Delta\theta= O(1/\bar{n})$ as having Heisenberg \textit{scaling}. However, this naming convention has not been universally adopted although, when we last checked, Caves is in favor of this declension \cite{tsang2012evading}.

The advantage of our currently proposed phase estimation scheme in this work is its experimental feasibility. Squeezed-vacuum generation in an optical parametric amplifier (OPA) of up to 11.5 dB of quadrature squeezing has been achieved experimentally \cite{PhysRevA.81.013814}. This in turn translates into a mean photon number
in both modes of the TMSV of up to about ten photons. Hence the parity, a measure of whether a state's photon number is even or odd, can be measured with existing
photon-number-resolving detectors such as transition edge sensors
\cite{a:TES:1998, irwin2005transition}, as well as with simpler homodyne methods \cite{plick2010parity}. However, a photon-number-resolving detector
does not provide a mean value of the parity signal after a single measurement,
which means that a phase measuring experiment must be repeated multiple times. 

Assuming a priori a flat distribution, our work applies Bayesian analysis to
the TMSV based phase-estimation scheme where photon-number-resolving detectors
are used to infer the parity signal. We use the parity signal at the output of the MZI to estimate the unknown phase $\theta$. Our analysis shows that, although phase estimation is biased near the phase origin and at $\pi/2$, there is an in-between interval where unbiased phase estimation is possible. In this interval, phase sensitivity saturates the Cram\'{e}r-Rao bound (CRB). Unlike the aforementioned phase estimation scheme, we achieve these results numerically with a finite number of parity measurements. Hence our result implies that parity is an optimal detection scheme for phase estimation with TMSV sources. 
%%*********Figure***********%
%%***************************%%
\section{Model}
%============================================
We consider a phase estimation scheme with a two-mode squeezed
vacuum (TMSV) input state which is commonly generated in unseeded optical parametric amplifiers. A TMSV state is ideally a superposition of twin Fock states
$\left\vert
  \psi_{\bar{n}}\right\rangle=\sum_{n=0}^{\infty}\sqrt{p_n\left(\bar{n}\right)}\left\vert
  n,n\right\rangle$, where the probability $p_{n}$ depends on the average number
of photons $\bar{n}$ in both modes of TMSV in the following way:
$p_n\left(\bar{n}\right)=(1-t_{\bar{n}})t^n_{\bar{n}}$ with
$t_{\bar{n}}=1/\left(1+2/\bar{n}\right)$ \cite{gerry2005introductory}. Propagation of the light through a MZI with linear phase accumulation imprints phase information on the state that is retrieved by measuring parity at the output of the MZI.

Parity based phase estimation was originally introduced in quantum optics by
Gerry \cite{PhysRevA.61.043811} and is based on the parity of the photon
number detected in the state at the output of the MZI. The expected value of the
parity signal $\langle\hat{\Pi}\rangle$ for a TMSV based phase estimation scheme, 
%%*********EQUATION***********%%
\begin{equation}
\langle\hat{\Pi}\rangle=\frac{1}{\sqrt{1+\bar{n}(\bar{n}+2)\sin^{2}\theta}},
\label{eq: Parity}
\end{equation} 
%%*****************************%%
was obtained in Ref.~\cite{PhysRevLett.104.103602}. It depends on the unknown
phase $\theta$ inside of the MZI and the mean photon number $\bar{n}$ in the
TMSV state used. Thus, knowing $\bar{n}$ at the input and
$\langle\hat{\Pi}\rangle$ at the output, the unknown phase $\theta$ can be
estimated. 

Each parity measurement returns either an even or odd outcome with
probabilities $P_{\rm e}$ and $P_{\rm o}$, respectively. Thus, there is uncertainty in
measuring the parity signal that is given by $\langle (\Delta
\hat{\Pi})^{2}\rangle=1-\langle\hat{\Pi}\rangle^{2}$, where the property $\hat{\Pi}^{2}=1$
has been used. Yet, in the limit of infinite measurement, the single-shot phase
sensitivity converges to $(\Delta \theta)^{2}=\langle (\Delta
\hat{\Pi})^{2}\rangle/(\frac{d \langle\hat{\Pi}\rangle}{d \theta}
)^{2}$. This implies that the uncertainty of phase estimation in the vicinity of $\theta=0$ is $(\Delta
\theta)^{2}=1/(\bar{n}^{2}+\bar{n})$ (that is, below the HL, as defined above). 

In practice, the parity measurement can be implemented with photon-number
resolving detectors. When using 100\% efficient photon-number-resolving detectors
one expects to detect $n$ photons with
probability $P(n)=\sum_{m>n/2}^{\infty}p_{m}(\bar{n})[d_{n-m,0}^{m}(\theta+\pi/2)]^{2}$, where
$p_{m}(\bar{n})$ is the probability of having a $|m,m\rangle$ state and
$d^{m}_{\mu,\nu}(\theta)$ is a rotation matrix element. Inferring the parity of
a state disregards the actual number of photons detected and focuses on
whether this number is even or odd. The probability of detecting an even
photon number is then $P_{\rm e}=\sum_{i=0}^{\infty}P(2i)$, where the summation can be evaluated to
%%*********EQUATION***********%%
\begin{equation}
  \label{eq:probs}
P_{\rm e}=\frac{1}{2}\left(1+\langle\hat{\Pi}\rangle\right)
\end{equation}
%%*****************************%%
since $P_{e}+P_{o}=1$ and the expectation value of a state's parity is $\langle\hat{\Pi}\rangle=P_{e}-P_{o}$.

Fig.~\ref{fig:ParityBuildUp} shows how an inferred parity signal (blue)
converges to its expected value of $\langle\hat{\Pi}\rangle=0.93$ (red) in  a
set of $M=10^{4}$ parity measurements. Here, we use the probability of an even
photon number, $P_{e}$ from Eq.~(\ref{eq:probs}), in order to numerically
generate one of many possible parity measurement records with an averaged input photon
number of $\bar{n}=3$ and an unknown phase of $\theta=0.1$. Due to the probabilistic nature of the scheme, repetition of this procedure with the same parameters results in a slightly different measurement record in each run. Hence, in any experiment there will always be an uncertainty in the inferred parity
signal that leads to an uncertainty in the estimation of the unknown phase $\theta$.

The statistics of the inferred variable $\phi$ is fully determined by the
statistics of the measured events $P_{e}$. Here, the $\phi$ is the estimate of the unknown phase $\theta$. A Bayesian approach to
interferometry provides an interval estimation and may be regarded as a
distribution of probability in the sense of degrees of likelihood \cite{0305-4470-31-2-015}. We
analyze a parity measurement record at the output of the MZI using Bayes'
theorem and obtain a probability for the unknown phase $\theta$ to be in the interval
$[\phi,\phi+d\phi]$ given all prior observations \emph{and a flat prior
distribution}. After each measurement, the probability density function (PDF)
is updated according to Bayes' theorem: $P(\phi|\rm output)
\propto P(\rm output|\phi)P_{\rm prior}(\phi)$, so that each consecutive observation modifies the PDF and improves phase estimation.
Hence, the update equation given by Bayes' theorem for an even or odd outcome is:
%%*********EQUATION***********%%
\begin{equation}
P(\phi|\{e,o\})\propto P(\{e,o\}|\phi)P(\phi).
\label{eq:Bays}
\end{equation}
%%*****************************%%
 Particular outcomes in our numerical model are independent, and thus 
the Bayesian PDF after $M$ runs with $m$ even outcomes is:
%%*********EQUATION***********%%
\begin{equation}
P(\phi|m)\propto P_{e}^{m}(\phi)P_{o}^{M-m}(\phi),
\label{eq:BaysM}
\end{equation}
%%*****************************%%
which provides an estimation interval with the PDF's maximum corresponding to the most
probable phase estimation.

\begin{figure}[t]
\centerline{\includegraphics[scale=0.6]{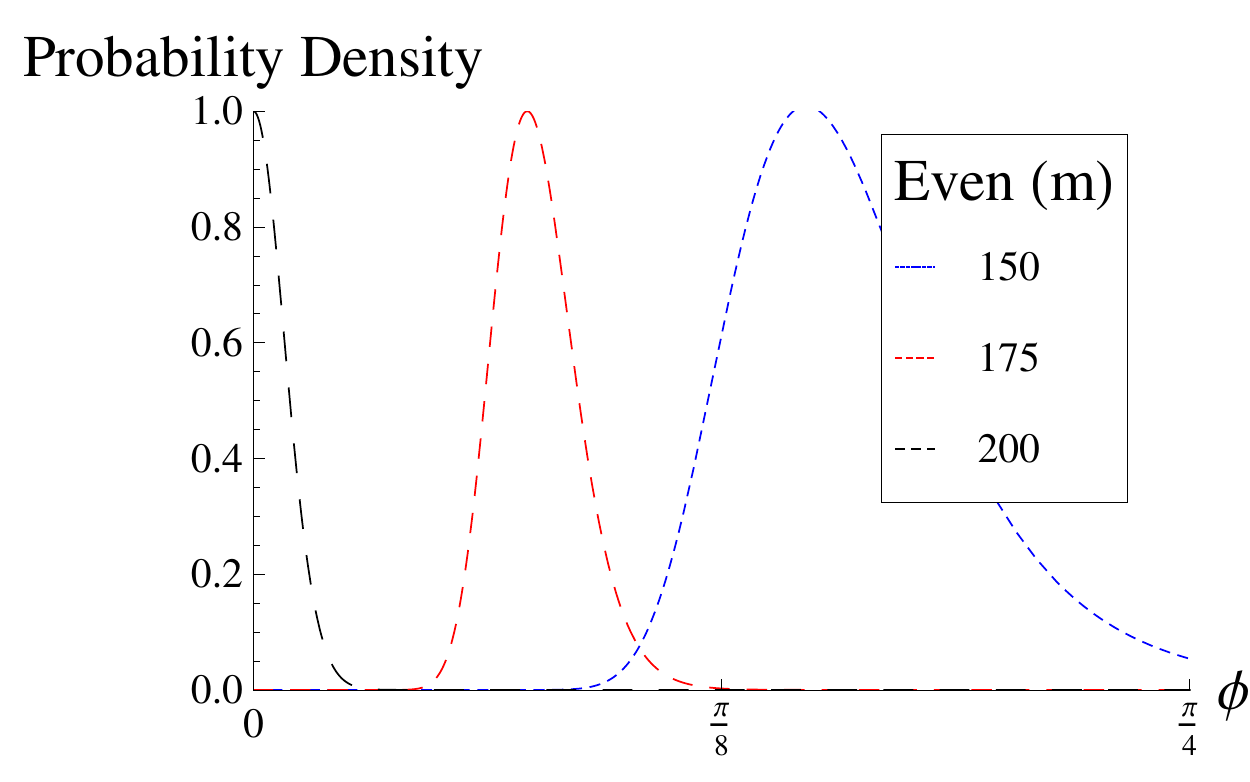}}
\caption{The Bayesian probability density function (PDF) from
  Eq.~(\ref{eq:BaysM}) is evaluated with mean photon number $\bar{n}=3$ and
  $M=200$ parity measurements for various numbers of even outcomes $m$. We
  rescaled the heights for easier comparison of their widths and maxima. As
  the number of even outcomes decreases from $m=M$ to $m=0.75M$, the maximum
  of the PDF shifts towards $\phi=\pi/2$ with a consequent broadening of the
  distribution that reduces the accuracy of phase estimation in this region.
}
\label{fig:ProbvsPhi}
\end{figure}

Fig.~\ref{fig:ProbvsPhi} presents an example of this update rule in the case
of $M=200$ runs with $\bar{n}=3$ input photons for different number of even
outcomes: $m={150,\ 175,\ 200}$. The heights of each have been rescaled for ease of comparison. One can see two effects that a reduced number of even
outcomes has on the PDF. The PDF becomes broader and shifts its maximum
towards $\phi=\pi/2$. This shift results in higher phase uncertainty and thus
reduction in the sensitivity of phase estimation. This result coincides with predictions of
Ref.~\cite{PhysRevLett.104.103602} that the best phase sensitivity is achieved
in the vicinity of $\theta=0$, where the parity of the output state is
predominantly even. 

%%*********Figure***********%%

%%%%%%%%%%%%%%%%%%%%%%%%%%%%
\section{Numerical results}
%%%%%%%%%%%%%%%%%%%%%%%%%%%%
%%%%%%%%%%%%%%%%%%%%%%%%%%%%

In contrast to having an assumed number of even outcomes as discussed in
Fig.~\ref{fig:ProbvsPhi}, the actual statistics of a finite-length measurement
record is governed by the following probability of even outcomes
%%*********EQUATION***********%%
\begin{equation}
  \label{eq:pe}
P_{e}=\frac{1}{2}\left(1+\frac{1}{\sqrt{1+\bar{n}(\bar{n}+2)\sin^{2}\theta}}\right),
\end{equation}
%%***************************%%
which comes directly from Eq.~(\ref{eq:probs}) and allows one to model phase
estimation with TMSV and parity detection numerically. Choosing the input
photon number $\bar{n}$ and an unknown phase $\theta$ turns the probability of
even outcomes $P_{e}$ into a number that we use to numerically generate a
measurement record of finite length $M$. From such a record, we determine a
number of even outcomes $m$ and obtain the Bayesian PDF using the update rule from Eq.~(\ref{eq:BaysM}). We base our single phase estimation on locating the
phase $\theta$ at the maximum of the Bayesian PDF.

There are four local maxima of the Bayesian PDF on a $2\pi$ interval due to the
periodicity of the parity signal and its invariance under inversion $\theta\to
-\theta$. Hence, unique phase values belong to the interval
$\theta\in\left(0,\pi/2\right]$ where the Bayesian PDF has a single maximum and
will be used for local phase estimation. In addition to phase estimation,
the Bayesian PDF provides an estimation interval that can be associated with the
width of a Bayesian PDF. In our case however, we adopt a different approach
to the calculation of such an interval so that we may better reflect the effects of
the measurement record length.

\begin{figure}[t]\centering
\includegraphics[scale=0.6]{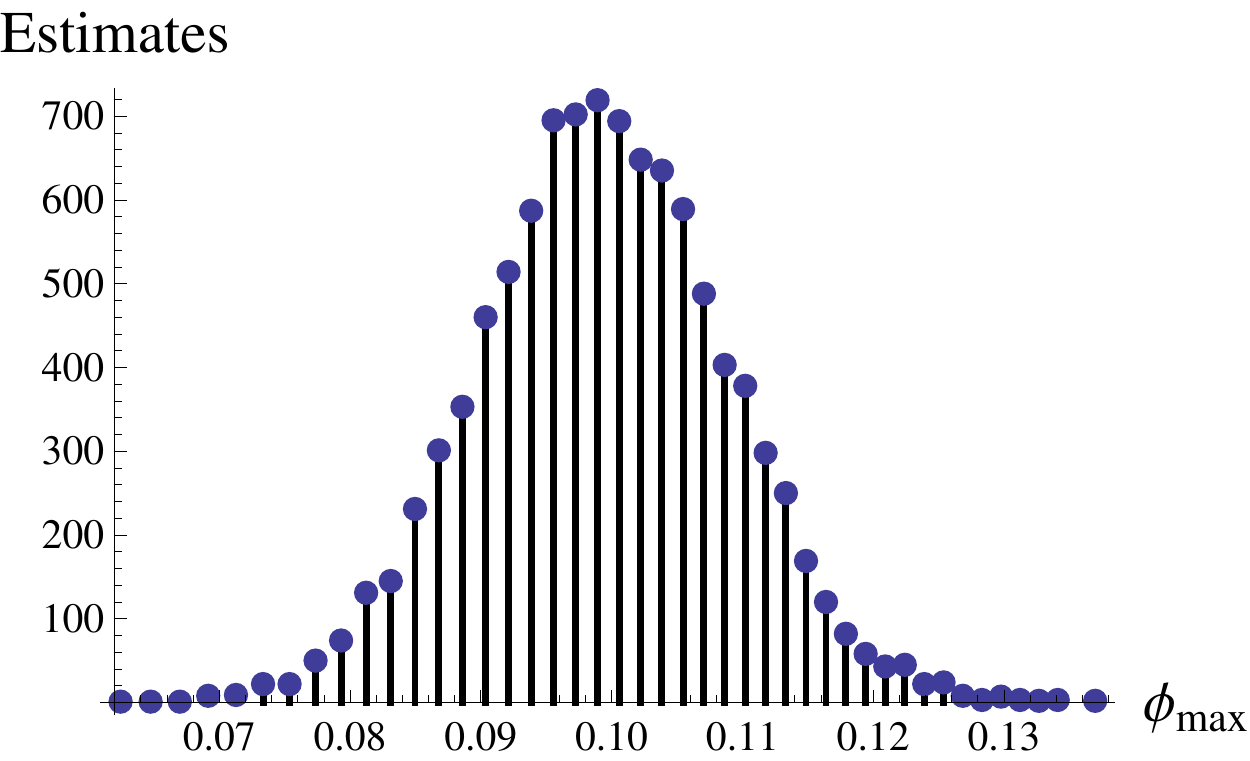}
\caption{Distribution of obtained phase estimations in a family of $N=10^{4}$
  parity measurement records that was numerically generated for $\theta=0.1$ and
  $\bar{n}=3$. The distribution shows uncertainty of a phase estimation $\phi$ due to a
  finite length of parity measurement record $M=10^{3}$. This distribution has
  a mean value of $\bar{\phi}=0.0997$ and a standard deviation of
  $\Delta\bar{\phi}=0.0094$ that characterizes the uncertainty of phase
  estimation after a finite number of measurements.}
\label{fig:MaxPhaseBarGraph}
\end{figure}

The most prominent effect of the record length on the phase estimation is the lack of reproducibility due to a finite deviation from the expected value of the parity signal. To compensate we repeat our experiment $N$ times. As an example, we have considered the family of $N=10^{4}$
parity measurement records generated for $\theta=0.1$, $\bar{n}=3$, and $M=10^{3}$. Fig.~\ref{fig:MaxPhaseBarGraph} shows the distribution of phase estimations, $\phi\in[0,\pi/2]$, for a measurement record in the family. There are three noteworthy points. First, possible phase estimations are discrete and sparse near $\phi=0$, which is associated with a number of odd outcomes in the measurement record. Second, the distribution has a mean value of $\bar{\phi}=0.0997$; hence, on average one would get a phase estimation with some bias, which is discussed in the following section. The final point, however, is that there is a spread of possible phase estimations with a standard deviation of $\Delta\bar{\phi}=0.0094$; hence, we will associate sensitivity of
the phase estimation with this standard deviation. 

%%*********Figure***********%%
%%***************************%%

%%%%%%%%%%%%%%%%%%%%%%%%%%%%
%%%%%%%%%%%%%%%%%%%%%%%%%%%%
\section{Bias and Phase Sensitivity}
%%%%%%%%%%%%%%%%%%%%%%%%%%%%
%%%%%%%%%%%%%%%%%%%%%%%%%%%%

As we ran our phase estimator with input intensity $\bar{n}=3$ for several unknown phases, $\theta\in[0,\pi/2]$, we found the presence
of bias, which is statistical favoritism that causes misleading results, defined as
\begin{equation}
{\rm Bias}=|\bar{\phi}-\theta|. 
\end{equation}

\begin{figure}
\centering
\subfloat[Part 1][]
{\includegraphics[width=3.2in]{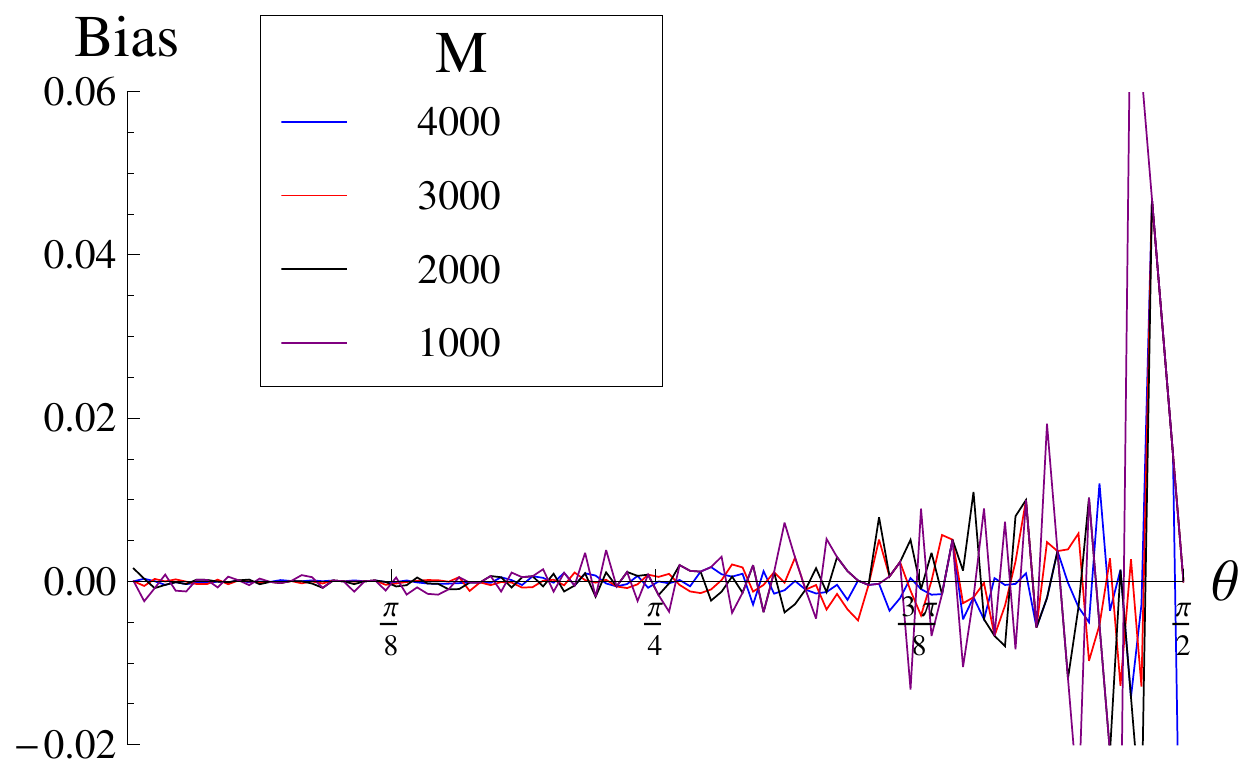} \label{fig:BiasVsTheta}}
\subfloat[Part 2][]
{\includegraphics[width=3.2in]{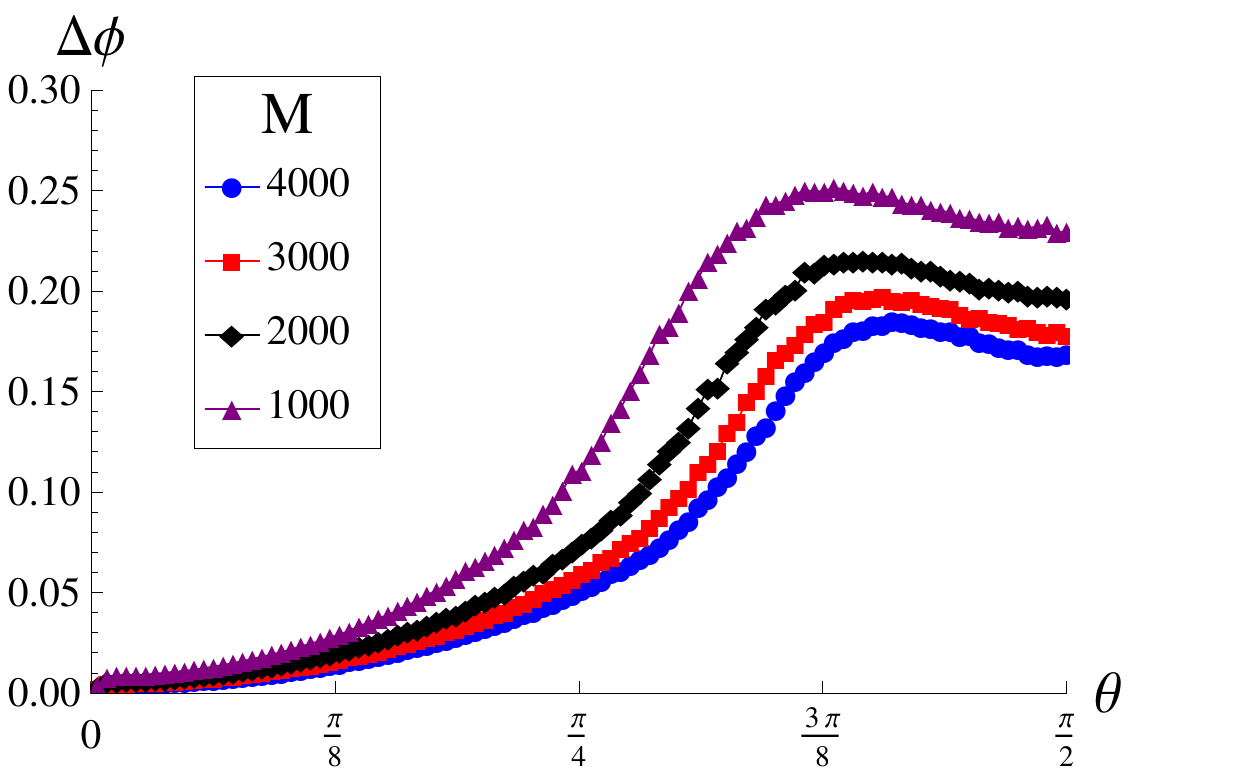} \label{fig:StDevVsTheta}}
\caption{(a) Bias of phase estimation for TMSV with $\bar{n}=3$ photons as a
  function of the unknown phase $\theta$. For a $N=10^{4}$ family of
  measurement records, we are increasing the length of the parity measurement records $M$. 
%Bias is negligible in the interval $\theta\in[0.02,1]$, and  thus provides us with an accurate phase estimation scheme.
(b) The standard deviation of phase estimation with an input photon number of $\bar{n}=3$ for increasing number of numerical runs $M$. We found that more parity measurements reduce the standard deviation. As $\theta$ increases towards $\pi/2$ the standard deviation becomes large. This explains why bias is high near $\theta=\pi/2$.}
\label{fig:FigureOne}
\end{figure}

Fig.~\ref{fig:BiasVsTheta} demonstrates the bias of our phase estimator for varying lengths of parity measurement records $M$. For all $M$ we see a high bias in estimating the unknown phase near $\theta=\pi/2$; however, longer measurement records $M$ maintain lower bias for a larger interval. 

%(*For an unknown phase $\theta\in[0,0.02]$, our phase estimation exhibits a large bias toward zero. This leaves an optimal phase interval of about $\theta\in[0.02, 1]$ that offers unbiased phase estimation for $\bar{n}=3$. 

The interval for unbiased phase estimation also depends on the input intensity $\bar{n}$. For a larger $\bar{n}=7$, our phase estimator shows that the minimum value of the unbiased phase interval remains relatively unchanged; however, the length of the interval reduces due to signal localization near the phase origin.
%Namely, the right boundary of the interval reduces from $\theta=0.6$ to $\theta=0.3$. (*This would depend on M*)

%%*********Figure***********%
%http://desk.stinkpot.org:8080/tricks/index.php/2006/05/place-multiple-images-on-one-line-in-latex-or-even-create-a-matrix-of-images/
%%***************************%%

The uncertainty of phase estimation $\Delta \phi$ quantifies the phase
sensitivity of the scheme. In the case of conventional phase estimation, with
coherent laser light and intensity difference measurement, phase sensitivity
is shot-noise limited, $\Delta \phi \geq1/\sqrt{M\bar{n}}$. This sensitivity
improves with increasing input intensity as well as the length of the
measurement record. However, the HL is,
$\Delta \phi \geq1/\sqrt{M\bar{n}^{2}}$, which has the same dependence as the  shot-noise on the length of the measurement record $M$ but faster dependence on the input photon number.

In the limit of a large number of parity measurements, phase estimation with
TMSV and parity detection is capable of beating the HL, as defined in the introduction. This is due to a greater photon
number variance in the TMSV state than is found in coherent states. The phase sensitivity for this scheme \cite{PhysRevLett.104.103602} is equal to the following:
%%*********EQUATION***********%%
\begin{equation}
  \label{eq:CRB}
  \Delta \phi=\frac{1+\bar{n}(2+\bar{n})\sin^2\theta}{\sqrt{M\bar{n}(2+\bar{n})}\cos\theta}.
\end{equation}
%%***************************%%
Hence, an optimum phase sensitivity, $\Delta
\phi=1/\sqrt{M\bar{n}(2+\bar{n})}$, obtained near $\theta=0$ is
sub-Heisenberg. More physically, this sensitivity approaches the CRB with parity detection.

In the case of finite-length measurement records, measuring parity with
photon-number-resolving detectors is biased near $\theta=0$ and $\pi/2$. Thus,
Bayesian phase estimation with photon-number resolving detectors can only be obtained in a low bias interval somewhere between $\theta=0$ and $\theta=\pi/2$. Fig.~\ref{fig:StDevVsTheta}
shows the standard deviation, $\Delta \phi$, of our phase
estimation scheme as a function of unknown phase $\theta$. Increasing the number of parity measurements in a phase estimate improves the phase sensitivity for all $\theta$. This is because phase uncertainty scales as
$\Delta\phi \propto c/\sqrt{M}$, where the proportionality constant $c$ depends on the unknown phase $\theta$ and input photon number $\bar{n}$, as per Eq.~\ref{eq:CRB}. By determining the value of $c$ the value of $\theta$ can be estimated. 

%Comparing the phase uncertainty to the bias we see that the values of bias are much less than the phase uncertainty meaning that we are obtaining accurate phase estimations \textbf{reword this sentence and place somewhere else}.

\begin{figure}
\centering
\subfloat[Part 1][Low end of unbiased interval.]
{\includegraphics[width=3.1in]{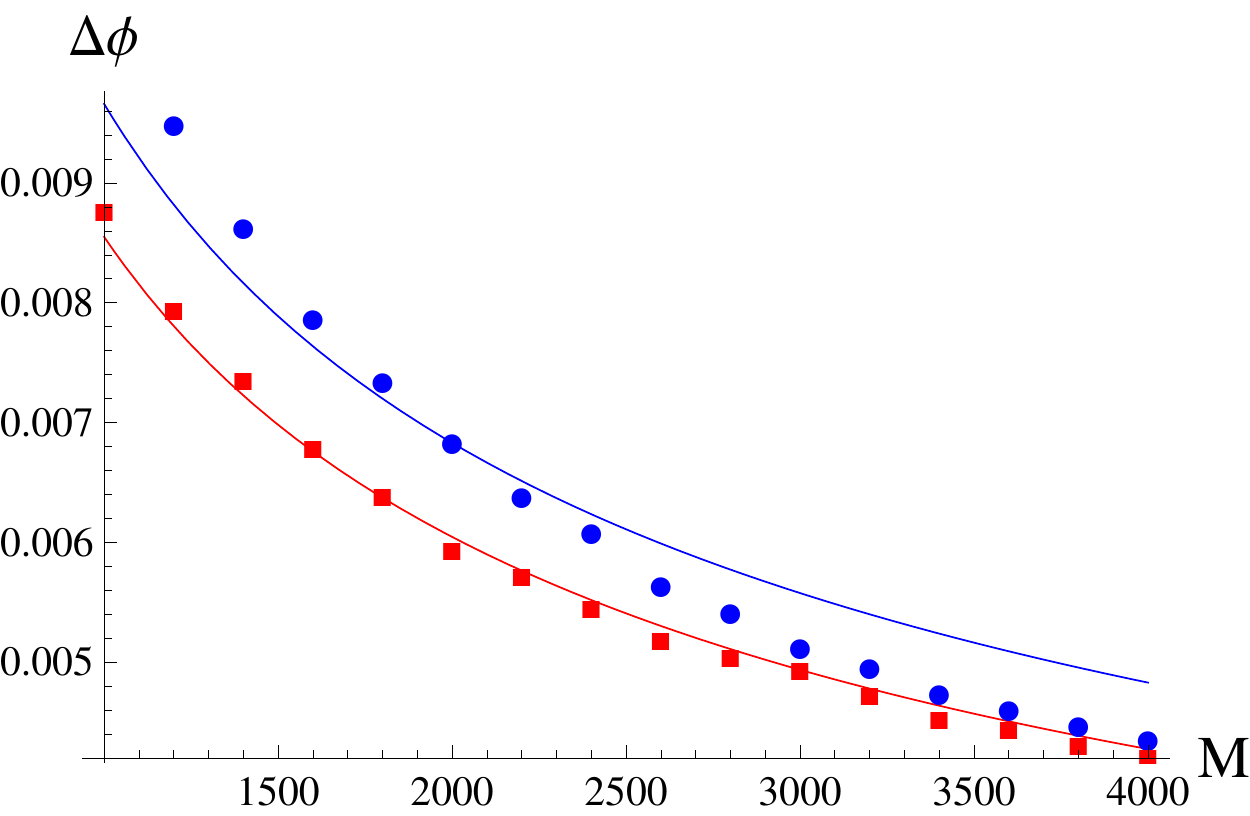} \label{fig:StDevVsTrials1}}
\subfloat[Part 2][High end of unbiased interval.]
{\includegraphics[width=3.1in]{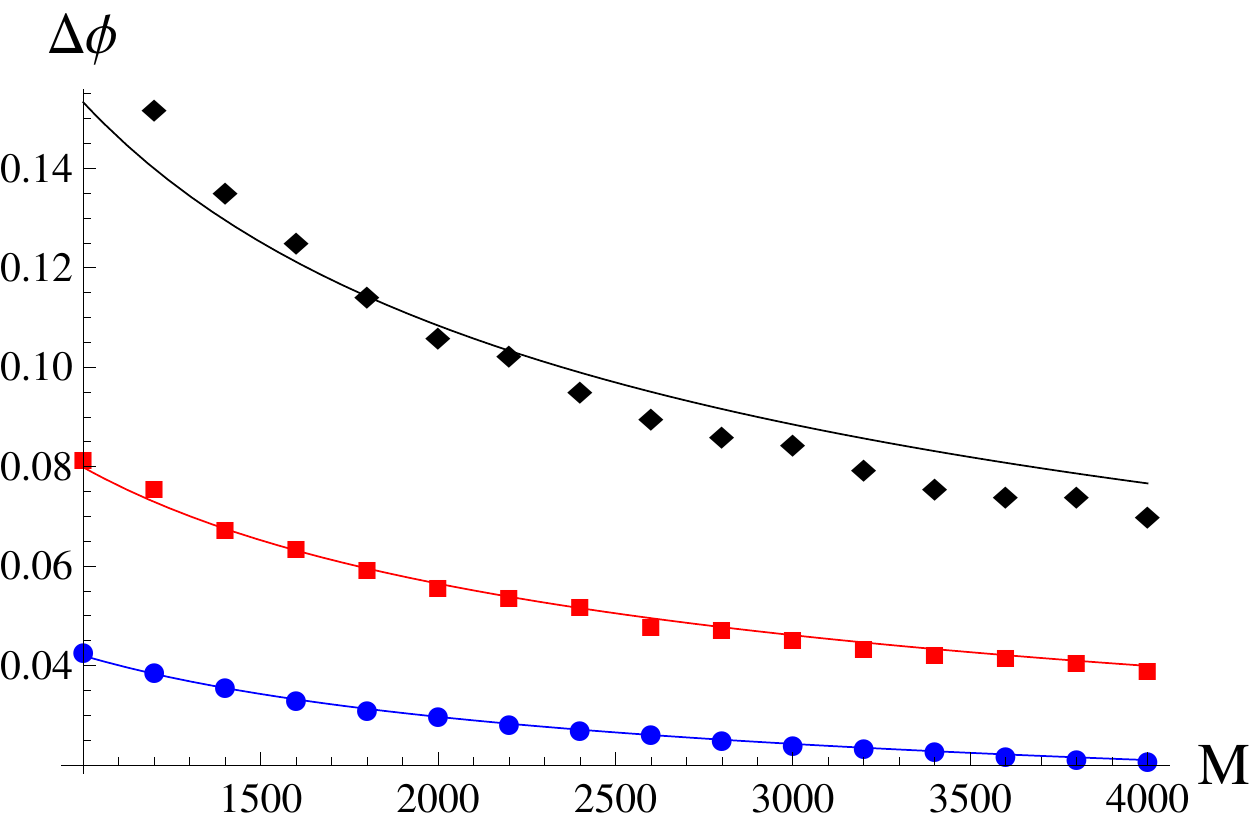} \label{fig:StDevVsTrials2}}
\caption{These figures show how the standard deviation changes for increasing numerical runs $M$ with $\bar{n}=3$ and varying unknown phases $\theta$. (a) The unknown phases are $\theta=0.02$ (blue circle) which exhibits poor scaling and $\theta=0.04$ (red square) which exhibits the expected $1/\sqrt{M}$ scaling. (b) The unknown phases are $\theta=0.9$ (blue circle), $\theta=1.1$ (red square), and $\theta=1.3$ (black diamond). When $\theta$ is greater than approximately $1$ the data deviates from the expected $1/\sqrt{M}$ scaling.}

\label{fig:FigureEnd}
\end{figure}

In Fig.~\ref{fig:FigureEnd} we calculate the standard deviation $\Delta 
\phi$ as a function of parity measurements $M$ for various unknown phases $\theta$ and fit them to their respective proportionality constant $c$. If the proportionality constant $c$ fits the data well one can obtain an accurate phase estimate. We can use this information to find a range of unknown phases $\theta$ that form an unbiased interval where accurate phase estimations can be made.

As an example for finding a value of $\theta$ outside of the unbiased interval, consider $\theta=0.02$. As we increase the number of numerical runs in a phase estimate, we find that the standard deviation remains roughly constant and cannot be fitted to a $c/\sqrt{M}$ function in order to determine the proportionality coefficient $c$, as shown on the blue curve in Fig.~\ref{fig:StDevVsTrials1}. This value of unknown phase is therefore in a region of high bias and an accurate phase estimation cannot be made.

%%*********Figure***********%
%http://desk.stinkpot.org:8080/tricks/index.php/2006/05/place-multiple-images-on-one-line-in-latex-or-even-create-a-matrix-of-images/

%%***************************%%

%%%%%%%%%%%%%%%%%%%%%%%%%%%%
%%%%%%%%%%%%%%%%%%%%%%%%%%%%
\section{Unbiased Phase Estimation}
%%%%%%%%%%%%%%%%%%%%%%%%%%%%
%%%%%%%%%%%%%%%%%%%%%%%%%%%%
Fig.~\ref{fig:StDevVsTrials1} shows how the standard deviation of phase estimation changes as a function of the number of parity measurement $M$ for $\bar{n}=3$, $N=10^{4}$, and unknown phases of $\theta=0.02$ (blue circle) and $\theta=0.04$ (red square). We find that the proportionality coefficients are $c_{\rm TMSV}=0.305$ and $0.270$ respectively. As we compare this proportionality coefficient with the SN limit, $c_{\rm SN}=0.58$, and the CRB, $c_{\rm CRB}=0.260$ and $0.265$ respectively, obtained from Eq.~(\ref{eq:CRB}), we see that the sensitivity is in close proximity to the CRB for $\theta=0.04$ but not for $\theta=0.02$. Futher examination shows that for values of unknown phase $0.02 < \theta \leq 0.04$ our scheme does not demonstrate the expected $1/\sqrt{M}$ dependence giving us the lower bound of our unbiased phase estimation interval to be $\theta \approx 0.04$. 

Similarly, Fig.~\ref{fig:StDevVsTrials2} presents this same information for unknown phases of $\theta=0.9$ (blue circle), $\theta=1.1$ (red square), and $\theta=1.3$ (black diamond). In these cases, we find that the proportionality coefficients are $c_{\rm TMSV}={4.85,\ 8.48,}$ and $9.49$ respectively. As we compare this proportionality coefficient with the SN limit, $c_{\rm SN}=0.58$, and the CRB, $c_{\rm CRB}={4.24,\ 7.35,}$ and $14.4$ respectively, we see that the sensitivity is in close proximity to the CRB for $\theta=0.9$. Futher examination shows that when $\theta$ is greater than approximately one our scheme does not demonstrate the expected $1/\sqrt{M}$ dependence which gives us the upper bound of our unbiased phase estimation interval to be $\theta\approx 0.9$. 

In the following, we focus on unknown phases in the interval suitable for unbiased phase estimation where our data does have the expected $1/\sqrt{M}$ dependence. Hence, we can compare the proportionality coefficient obtained from the fitting procedure with a value expected from the CRB as well as from the SN limit.

\begin{figure}
\centering
\subfloat[Part 1][]
{\includegraphics[width=3.2in]{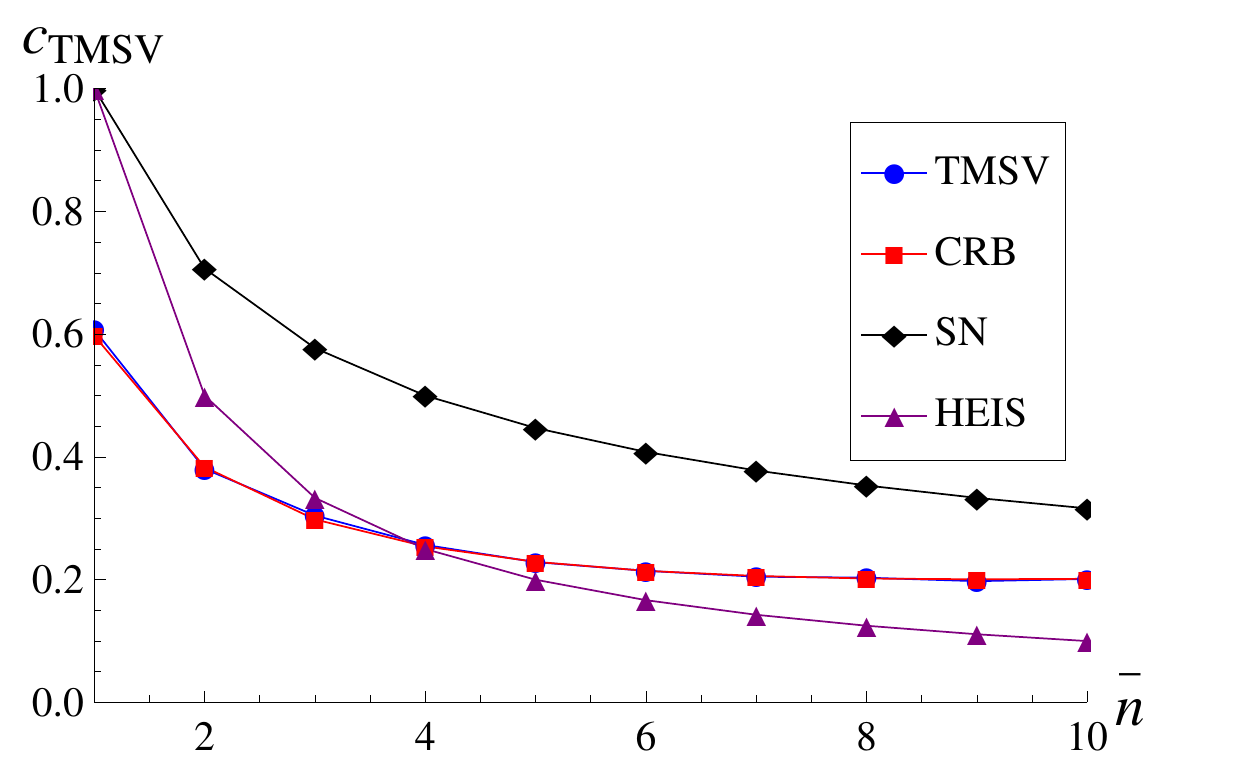} \label{fig:CvsN1}}
\subfloat[Part 2][]
{\includegraphics[width=3.2in]{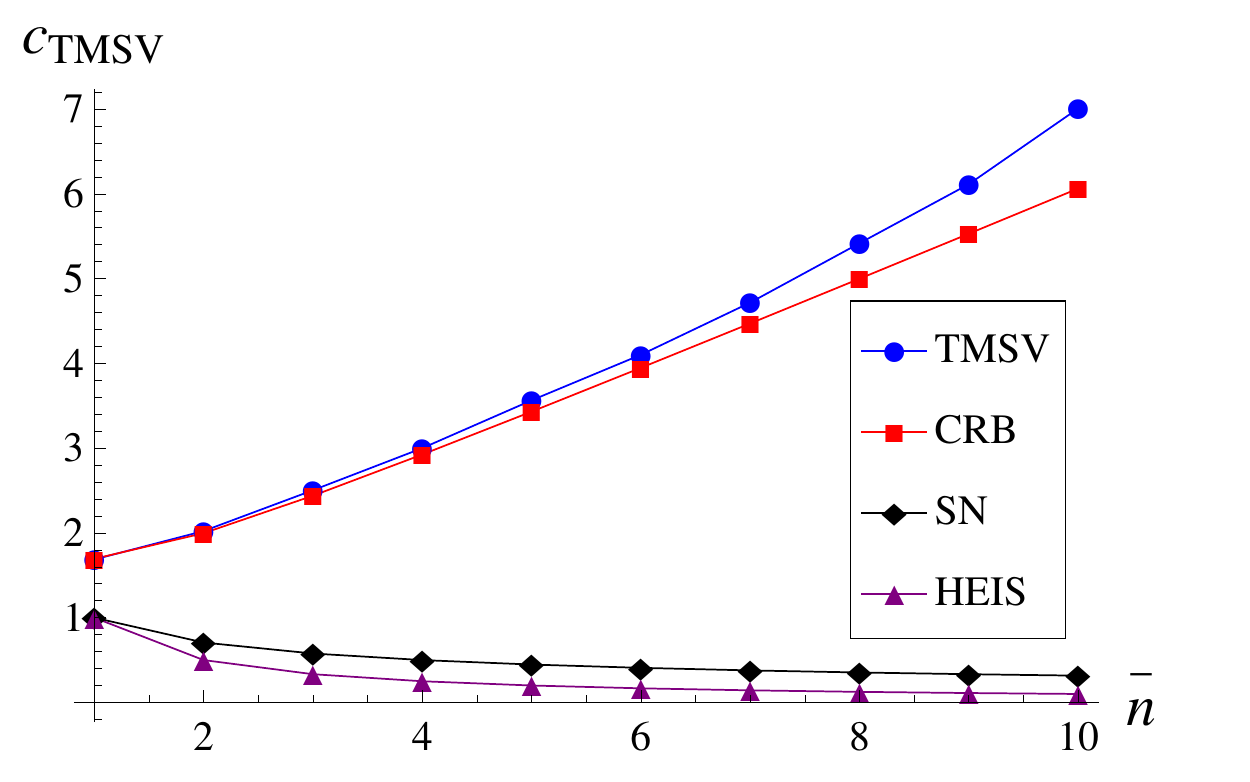} \label{fig:CvsN2}}
\caption{The constant of proportionality, $c_{\rm TMSV}$, versus the number of photons in the input state. The SN limit in black and the Heisenberg limit (HL) in purple are present for comparison. (a) For $\theta=0.1$ the demonstrated sensitivity is in close proximity to the CRB in red.
(b) For $\theta=0.7$ the demonstrated sensitivity is in close proximity to the CRB in red for low $\bar{n}$ but begins to deviate away for larger $\bar{n}$.}
\label{fig:FigureEnd2}
\end{figure}

For a given unknown phase, the sensitivity of phase estimation can be
improved by varying the mean photon number in the input TMSV state. This is analyzed for $\theta=0.1$ and $\theta=0.7$ and the results are presented in Fig.~\ref{fig:FigureEnd2}, where the phase sensitivity is characterized by $c_{\rm TMSV}$. This removes the dependence on $M$ and focuses on the input intensity. For unknown phases $\theta\in[0.04,0.9]$, phase sensitivity is in close proximity to the CRB. 
%%*********Figure***********%
%http://desk.stinkpot.org:8080/tricks/index.php/2006/05/place-multiple-images-on-one-line-in-latex-or-even-create-a-matrix-of-images/

%%***************************%%

The best sensitivity of our phase estimation protocol is expected near the phase origin. Performing phase estimation in the unbiased interval
offers significant improvement for phase sensitivity. With an unknown phase of $\theta=0.1$ shown in Fig.~\ref{fig:CvsN1}, one can see that phase sensitivity of our scheme saturates the CRB and beats the SNL for an extended range of input intensities. For phase estimation in the unbiased interval but farther away from the phase origin we still saturate the CRB for small input intensities but begin to deviate from this bound for higher input intensities. Fig.~\ref{fig:CvsN2} demonstrates this with an unknown phase of $\theta=0.7$.

Thus, parity measurement with photon-number-resolving detectors is at the limiting performance in the unbiased interval of $\theta\in[0.04,0.9]$ for our scheme and
could be used experimentally to estimate unknown phases at the CRB with a finite number of resources.
%%***************************%%
%%%%%%%%%%%%%%%%%%%%%%%%%%%%
%%%%%%%%%%%%%%%%%%%%%%%%%%%%
\section{Conclusion}
%%%%%%%%%%%%%%%%%%%%%%%%%%%%
%%%%%%%%%%%%%%%%%%%%%%%%%%%%
Phase estimation protocols benefit from
photon number-resolving-detectors by inferring the
parity of the state. Here, we considered a particular phase estimation
protocol that is based on two-mode squeezed-vacuum (TMSV) input and parity detection at the output of a Mach-Zehnder interferometer (MZI). The use of photon-number-resolving detectors means that measurements must be repeated multiple times with Bayesian analysis applied to all outcomes. 

Our scheme shows that phase sensitivity
saturates the Cram\'{e}r-Rao bound and beats the Heisenberg limit (as defined for this work) with the use of a \emph{finite number} of experimental runs. We discovered that our maximum-likelihood estimator is biased near the phase origin where the best sensitivity is
expected. As a result, the standard deviation of the estimated values does not
reduce with increasing number of phase estimations but stays constant. However, values of the unknown phase in our unbiased interval $\theta\in[0.04,0.9]$ allow 
for unbiased phase estimation. Phase sensitivity behaves as expected and remains in close
proximity to the Cram\'{e}r-Rao bound. Consequently, the phase sensitivity is optimal and remains sub-shot-noise limited for a broad range of input intensities as long as phase estimation is performed near the origin.

%%%%%%%%%%%%%%%%%%%%%%%%%%%%
%%%%%%%%%%%%%%%%%%%%%%%%%%%%
\section*{Acknowledgments}
%%%%%%%%%%%%%%%%%%%%%%%%%%%%
%%%%%%%%%%%%%%%%%%%%%%%%%%%%
We would like to acknowledge the Air Force Office of Scientific Research and the National Science Foundation for support. We would also like to thank Tom Scholz for helpful discussions.

\bibliographystyle{osajnl}
\bibliography{References}

\end{document}